\documentclass[fleqn,10pt]{wlscirep}
\usepackage[utf8]{inputenc}
\usepackage[T1]{fontenc}
\usepackage{float}
\usepackage{amsmath}
\usepackage{amssymb}
\usepackage{pgfplots}
\usepackage{lineno}
\usepackage{setspace}
\usepackage{geometry} 
\newcommand{\bibcommenthead}[1]{} 
\title{Examples of Atoms Absorbing Photon via Schrödinger Equation and Vacuum Fluctuations}
\author[1,*]{Yongjun Zhang}
\affil[1]{Science College, Liaoning Technical University, Fuxin, 123000, China}
\affil[*]{yong.j.zhang@gmail.com}
\begin{abstract}
The absorption of photons by atoms encompasses fundamental quantum mechanical aspects, particularly the emergence of randomness to account for the inherent unpredictability in absorption outcomes. We demonstrate that vacuum fluctuations can be the origin of this randomness. An illustrative example of this is the absorption of a single photon by two symmetrically arranged atoms. In the absence of a mechanism to introduce randomness, the Schrödinger equation alone governs the time evolution of the process until an entangled state of the two atoms emerges. This entangled state consists of two components: one in which the first atom is excited by the photon while the second remains in the ground state, and another in which the first atom remains in the ground state while the second is excited by the photon. These components form a superposition state characterized by an unbreakable symmetry in the absence of external influences. Consequently, the absorption process remains incomplete. When vacuum fluctuations come into play, they can induce fluctuations in the weights of these components, akin to Brownian motion. Over time, one component diminishes, thereby breaking the entanglement between the two atoms and allowing the photon absorption process to conclude. The remaining component ultimately determines which atom completes the photon absorption. Similar studies involving different numbers of atoms can be conducted. Vacuum fluctuations not only introduce randomness but also have the potential to give rise to the Born rule in this context. Furthermore, the Casimir effect, which is closely tied to vacuum fluctuations, presents a promising experimental avenue for validating this mechanism.
\end{abstract}

\begin{document}
\renewcommand{\hbar}{\mathchar'26\mkern-9mu h}

\flushbottom
\maketitle

\newgeometry{left=1in, right=3in, top=1in, bottom=1in}


\section{Introduction}
Quantum optics probes the intricate interactions between light and matter, surpassing the realm of classical optics to examine the behavior of individual photons and their interactions with atoms and subatomic particles~\cite{10.1017/CBO9780511813993}. A fundamental aspect in this domain is the absorption of photons by atoms, a core quantum-mechanical process characterized by an atom transitioning from a lower to a higher energy state, the excited state, upon photon absorption. Studying this process offers crucial insights into quantum mechanics, particularly in unraveling four interconnected perspectives that remain less understood.

First, there is a challenge in aligning quantum unitary time evolution~\cite{10.1007/978-1-4757-0576-8} with the discrete eigenstates of atoms. Consider a hydrogen atom transitioning from its ground state to an excited state due to photon absorption. In a scenario without discrete states, unitary time evolution could seamlessly map the initial state to the final state. However, the inherent discreteness of atomic eigenstates complicates this mapping. According to measure theory~\cite{zotero-937}, such a direct mapping would effectively have a measure of zero. Consequently, to enable a non-zero transition probability, the rigid framework of unitary time evolution must be relaxed. This relaxation should allow multiple initial states to converge to the same final state, inherently leading to some loss of information from the initial state and the subsequent emergence of randomness.

Second, the nature of randomness in quantum mechanics remains enigmatic. Einstein's famous dictum, "God does not play dice," challenges the notion of intrinsic randomness and resonates through various interpretations of quantum phenomena. The Copenhagen Interpretation~\cite{10.1038/121580a0,10.1007/BF01397280} attributes randomness to wave function collapse according to the Born rule~\cite{10.1007/BF01397184}. However, this interpretation has been contested due to its implications of observer-dependency and perceived incompleteness of reality~\cite{10.1007/BF01491891,10.1103/PhysRev.47.777a,10.1103/PhysicsPhysiqueFizika.1.195a}. Alternative approaches include the deterministic pilot wave theory, which posits hidden variables and a guiding wave~\cite{10.1051/jphysrad:0192700805022500,10.1103/PhysRev.85.166,zotero-956}, and the many-worlds interpretation~\cite{10.1103/RevModPhys.29.454,saunders2010many}, which suggests a branching universe for each quantum event, rendering randomness as emergent. Decoherence theory, on the other hand, deals with randomness at the statistical level of quantum ensembles~\cite{10.1007/BF00708656,10.1103/PhysRevD.24.1516,joos2003decoherence,10.1103/RevModPhys.75.715,10.1103/RevModPhys.76.1267,10.1016/j.physrep.2019.10.001}. The Ghirardi-Rimini-Weber (GRW) model~\cite{10.1103/PhysRevD.34.470,10.1103/PhysRevA.40.1165,10.1016/S0370-15730300103-0,10.1007/978-3-030-31146-9_1} and its successor, the Continuous Spontaneous Localization (CSL) model~\cite{10.1103/PhysRevA.39.2277,10.1103/PhysRevA.42.78,10.1103/RevModPhys.85.471,10.1103/PhysRevD.94.124036}, introduce spontaneous collapse into the quantum framework to account for state vector reduction, but the source of underlying randomness remains unspecified.

Third, research on ensembles may not fully elucidate the dynamics of individual quantum systems, underscoring the need for distinct approaches at these two analytical levels. An ensemble-based approach is often indispensable and more feasible, particularly when detailed knowledge of the underlying dynamics is sparse. A prime example is Einstein's investigation of Einstein coefficients~\cite{Einstein1917} through thermodynamic considerations, a domain intrinsically tied to the concept of ensembles. However, it is crucial to recognize that ensemble studies are not a replacement for delving into the intricacies of individual quantum systems. This becomes especially apparent in the context of the Born rule: while its application in ensemble scenarios is often direct and taken for granted, in the case of single systems, the rule's relevance is dictated by the dynamics of the system itself, which define all potential outcomes and their probabilities. Therefore, the Born rule must naturally arise from these dynamics. Elements of dynamics involving randomness, which might be overlooked in ensemble-focused theories, are essential in the analysis of individual systems. In the context of the CSL model, Pearle's use of the gambler’s ruin problem as an analogy has facilitated the derivation of the Born rule for individual quantum systems~\cite{10.1007/BF00726850,shimony_honor_1997}.

Finally, determining the exact nature of photon absorption—whether it is continuous, instantaneous, or a hybrid process starting continuously and completing instantaneously—remains an open question. Bohr's model~\cite{Bohr1913-BOHOTC} depicts electron transitions in hydrogen atoms as instantaneous, a view echoed by the Copenhagen interpretation, especially in the context of treating photon absorption as a quantum measurement. This perspective aligns well with the randomness introduced by the Born rule. On the other hand, the CSL model proposes a continuous process for introducing randomness~\cite{10.1103/PhysRevA.39.2277,10.1103/PhysRevA.42.78}, yet it contends with the 'tail problem,' where remnants of the wave function persist post-collapse~\cite{shimony_honor_1997}. Furthermore, the derivation of Einstein coefficients within quantum mechanics implies a continuous process~\cite{zotero-941}. Extending this concept to quantum electrodynamics~\cite{zotero-932} suggests that vacuum fluctuations play a role in atom-photon interactions and may trigger spontaneous emission~\cite{10.1103/PhysRevA.11.814,10.1119/1.13886}. These fluctuations, arising from Heisenberg's uncertainty principle~\cite{10.1007/BF01397280}, occur because a quantum field’s amplitude and rate of change cannot simultaneously be zero, even in a vacuum. The Casimir effect provides empirical evidence for these vacuum fluctuations~\cite{10.1103/PhysRev.73.360,10.1103/PhysRevLett.78.5}.

This paper investigates the absorption of a single photon by atoms, based on two hypotheses: the absorption process is continuous, and its inherent randomness arises from vacuum fluctuations. Our focus is primarily on individual systems, as opposed to ensembles. We begin by studying an example of two atoms absorbing a single photon. Next, we examine the role of vacuum fluctuations and draw an analogy with Brownian motion. Following this, we propose methods to validate the mechanism experimentally. Subsequently, we analyze two additional examples before discussing the interplay of information and randomness.


\section{Two Atoms Absorbing One Photon}
\begin{figure}[h]
    \centering
    \includegraphics[width=7cm]{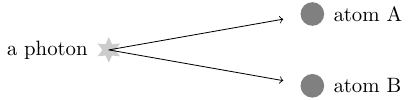}
    \caption{Two atoms interacting with a single photon. The photon initially interacts symmetrically with both atoms but ultimately is absorbed by only one, breaking the symmetry and introducing randomness.}
    \label{atoms}
\end{figure}

We consider a scenario depicted in Fig.~\ref{atoms}, where a single photon interacts with two identical atoms that are initially in their ground states. For simplicity, we assume that each atom has only two energy levels: the ground state and an excited state, with an energy difference of \( \Delta E \). If the frequency \( \nu \) of the photon is off-resonance---i.e., \( h \nu \neq \Delta E \), where \( h \) is Planck's constant---the atoms will not absorb the photon. Instead, the photon will be diffracted, leading to an interference pattern. In an extended atomic lattice, this pattern would correspond to Bragg's law~\cite{10.1098/rspa.1913.0040}.

The system is initially symmetric, with the photon's wave function interacting equally with both atoms. However, this symmetry is broken upon photon absorption, as one atom becomes excited while the other remains in the ground state. This symmetry breaking introduces randomness.

An instantaneous process would imply infinite changes in energy and momentum, which are physically unrealistic. Therefore, at least in the initial stages of absorption, the process is continuous. This process is governed by the Schrödinger equation:
\begin{equation}
i\ \hbar\frac{d}{dt}|\Psi_{\text{I}}\rangle = H_{\text{I}}|\Psi_{\text{I}}\rangle,
\end{equation}
where $|\Psi_{\text{I}}\rangle$ represents the overall wave function of the system, including both atoms and the photon. Here, $H_{\text{I}}$ is the Hamiltonian, and the subscript $\text{I}$ indicates that we are working within the interaction picture.

The initial symmetry may be maintained if both atoms partially absorb and then re-emit the photon, similar to a double-slit experiment. This would result in an observable interference pattern~\cite{10.1103/PhysRevLett.70.2359}. However, to absorb the photon, the symmetry must be broken. The deterministic nature of the Schrödinger equation alone is insufficient to achieve this break in symmetry. When the equation is unable to further evolve the system, it essentially "freezes", leading to:
\begin{equation}
    i\ \hbar\frac{d}{dt}|\Psi_{\text{I}}\rangle = 0.
\end{equation}
At this point, the overall wave function takes the form:
\begin{equation}\label{3}
|\Psi_{\text{I}}\rangle = \frac{1}{\sqrt{2}}|1\rangle_A|0\rangle_B + \frac{1}{\sqrt{2}}|0\rangle_A|1\rangle_B,
\end{equation}
where $|0\rangle$ and $|1\rangle$ denote the ground and excited states of an atom, respectively.
The two atoms are entangled.

\section{Superposition of Two Discrete Eigenstates}
We can rewrite Eq. (\ref{3}) as follows:
\begin{equation}\label{4}
|\Psi_{\rm I}\rangle = \sqrt{2}\left(\frac{1}{\sqrt{2}}\vert 0\rangle+\frac{1}{\sqrt{2}}\vert 1\rangle\right)_A\left(\frac{1}{\sqrt{2}}\vert 0\rangle+\frac{1}{\sqrt{2}}\vert 1\rangle\right)_B-\frac{1}{\sqrt{2}}\vert 0\rangle_A\vert 0\rangle_B-\frac{1}{\sqrt{2}}\vert 1\rangle_A\vert 1\rangle_B.
\end{equation}
The coefficient $\sqrt{2}$ in the first term arises due to the non-orthogonality of the three terms. Note that the first term is unique and cannot be represented by any combination of the remaining terms. This term captures a unique feature of the system: both atom \( A \) and atom \( B \) are at least partially in the intermediate state $\frac{1}{\sqrt{2}}\left(\vert 0\rangle+\vert 1\rangle\right)$. Additionally, while the components $\vert 0\rangle_A\vert 0\rangle_B$ and $\vert 1\rangle_A\vert 1\rangle_B$ may not satisfy the laws of energy conservation, they are permitted to appear for a very short time due to Heisenberg's uncertainty principle~\cite{10.1007/BF01397280}. 

Despite the entanglement between atoms \( A \) and B, each atom can still partially behave as if it were an isolated particle in a superposition state. This phenomenon can be further illustrated by the hydrogen atom, a two-body system in which the electron and proton are positionally entangled. Nonetheless, the electron can still be described by a wave function, such as \( \psi_{1,0,0}(r,\theta,\phi) \), as if it were an isolated particle, albeit in a superposition state.

Electrons in atoms occupy discrete eigenstates, resulting in stable wave functions over time scales shorter than the lifetime of spontaneous emission. For example, in a hydrogen atom's 2p state, the electron's density distribution remains time-independent over durations much shorter than the 2p state's nanosecond lifetime.

In contrast, when considering a superposition of discrete eigenstates, the wave function becomes dynamic. Let's consider atoms \( A \) and \( B \) as hydrogen atoms, with their eigenstates represented as follows:
\begin{equation}
|0\rangle = \Psi_{1,0,0}(r,\theta,\phi,t), \quad |1\rangle = \Psi_{2,1,1}(r,\theta,\phi,t),
\end{equation}
with corresponding energy eigenvalues \( E_{1,0,0} \) and \( E_{2,1,1} \), respectively.
When atom \( A \) is in the state \( \frac{1}{\sqrt{2}} (|0\rangle + |1\rangle) \), the wave function is given by:
\begin{equation}\label{Psi}
\Psi_A(r,\theta,\phi,t) = \frac{1}{\sqrt{2}} \Psi_{1,0,0}(r,\theta,\phi,t) + \frac{1}{\sqrt{2}} \Psi_{2,1,1}(r,\theta,\phi,t).
\end{equation}
Consequently, the time-dependent density distribution is:
\begin{equation}\label{f2}
|\Psi_A(r,\theta,\phi,t)|^2 = f_1(r,\theta) + f_2(r,\theta) \cos(\phi - \omega t),
\end{equation}
where \( \omega = \Delta E / \hbar \), or more specifically, \( \omega = (E_{2,1,1} - E_{1,0,0}) / \hbar \), \( \hbar \) is the reduced Planck's constant. This density distribution rotates with an angular frequency \( \omega \), potentially leading to an emission of electromagnetic waves or interactions with existing fields. In a physical vacuum, electromagnetic waves may arise due to vacuum fluctuations and may interact with this state. In addition, cosmic microwave background radiation~\cite{10.1086/148307} may also play a role.

\section{Wave Function Collapse Induced by Vacuum Fluctuations}

When atom \( A \), in the state described by Eq.~(\ref{Psi}), interacts with a vacuum fluctuation, its wave function undergoes a change:
\begin{equation}
\Psi_A(r,\theta,\phi,t) \xrightarrow[\text{fluctuation}]{\text{a vacuum}} \sin\phi \Psi_{1,0,0}(r,\theta,\phi,t) + \cos\phi \Psi_{2,1,1}(r,\theta,\phi,t).
\end{equation}
A phase difference between the two terms may also exist, but it should not play a major role in this study.
The energy change from \( \frac{1}{2}(E_{1,0,0} + E_{2,1,1}) \) to \( \sin^2\phi E_{1,0,0} + \cos^2\phi E_{2,1,1} \) necessitates a corresponding change in atom \( B \) to conserve energy:
\begin{equation}
\Psi_B(r,\theta,\phi,t) \xrightarrow[\text{fluctuation}]{\text{same vacuum}} \cos\phi \Psi_{1,0,0}(r,\theta,\phi,t) + \sin\phi \Psi_{2,1,1}(r,\theta,\phi,t).
\end{equation}

This instantaneous correlation between possibly widely separated atoms is a hallmark of quantum entanglement, initially proposed by Einstein, Podolsky, and Rosen (EPR)\cite{10.1103/PhysRev.47.777a} and further elaborated by Bell\cite{10.1103/PhysicsPhysiqueFizika.1.195a}. The entanglement persists from the moment the photon starts to interact with the atoms until one of the atoms absorbs the photon completely, thereby linking the photon absorption process with both the creation and eventual destruction of entanglement between the atoms. 

Accordingly, the overall wave function evolves as follows:
\begin{equation}\label{co}
|\Psi_{\text{I}}\rangle \xrightarrow[\text{fluctuation}]{\text{vacuum}} \sin\theta|1\rangle_A|0\rangle_B + \cos\theta|0\rangle_A|1\rangle_B,
\end{equation}
where \( \theta = \arctan\cot^2\phi \) and
any possible phase difference between the two terms has been omitted. This expression ensures the conservation of energy. However, the Heisenberg uncertainty principle permits brief departures from the usual rule of energy conservation. This means that the above expression can be temporarily deviated from, allowing the two atoms to interact with vacuum fluctuations independently for a short time. Eventually, these independent interactions reconcile to ensure energy conservation over a longer timescale.

The entanglement between the two atoms suggests the presence of faster-than-light correlations. Interestingly, this faster-than-light feature is also present within each individual atom. As described by Eqs.~(\ref{Psi}) and (\ref{f2}), the electron cloud extends infinitely, giving the appearance of superluminal rotation in its outer regions. However, this does not violate causality, as the electron cloud represents a single quantized entity rather than a multi-particle system. 
Causality is preserved as long as the wave function's propagation speed to regions where it previously had zero amplitude does not exceed the speed of light from the nearest region with non-zero amplitude.

Another feature is that vacuum fluctuations induce a smooth process to finish the photon absorption. As the system approaches an eigenstate, the factor \( f_2(r,\theta) \) in Eq.~(\ref{f2}) decreases, weakening the interaction with vacuum fluctuations. This results in a gradual slowing down of the dynamic process, ensuring that the dynamic process is continuous even at its conclusion. This phenomenon can be further studied using an analogy with Brownian motion.

Like the CSL model~\cite{shimony_honor_1997}, our example faces a "tail problem," but in the temporal dimension, characterized by indefinite photon absorption duration. It's worth noting that some processes in physics, such as the formation of a hydrogen atom, inherently demand an infinite duration to comply with causality. For instance, an electron transitioning from a confined state to an eigenstate, such as $\Psi_{1,0,0}(r,\theta,\phi,t)$, which extends to infinity, must take an infinite amount of time. In light of this, the temporal tail problem in our example doesn't pose a fundamental issue. Its experimental impact is minimal, as the effect diminishes exponentially over time, rendering it practically negligible. 
In fact, this exponential thinning is so rapid that, on the scale of physical experiments, the entire process appears to occur almost instantaneously.

\section{Brownian Motion Analogy for Disentanglement}
For the example illustrated in Fig.~\ref{atoms}, the disentanglement begins when the overall wave function is in the state described by Eq.~(\ref{3}). In a more general scenario where the two atoms are not arranged symmetrically, the system's wave function, when it is stuck, takes the following form:
\begin{equation}\label{psi}
|\Psi_{\text{I}}\rangle = C_1|1\rangle_A|0\rangle_B + C_2|0\rangle_A|1\rangle_B,
\end{equation}
where \(C_1\) and \(C_2\) are complex probability amplitudes that satisfy \(|C_1|^2 + |C_2|^2 = 1\). We define a parameter \(x\) such that
\begin{equation}
|C_1|^2 = x, \quad |C_2|^2 = 1-x.
\end{equation}

\begin{figure}[h]\centering
        \includegraphics[width=7cm]{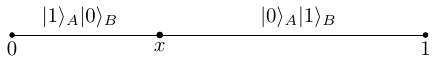}
\caption{
Brownian motion on a line of unit length, where a particle starts at position \(x\). The probability of the particle reaching endpoint 1 is \(x\), while the probability of the particle reaching endpoint 0 is \(1-x\).
}\label{1point}
\end{figure}

The disentanglement process in this general case can be likened to a particle undergoing continuous Brownian motion \cite{10.1063/1.1895505} on a unit-length line, starting at position \(x\), as depicted in Fig. \ref{1point}. The particle reaching the endpoints 0 and 1 corresponds to the disentanglement outcomes \(|0\rangle_A|1\rangle_B\) and \(|1\rangle_A|0\rangle_B\), respectively. To model the slowing down of the disentanglement near eigenstates, we introduce a diffusion coefficient \cite{10.1016/j.physa.2018.10.017} \(D(x)=x(1-x)\). Consequently, the diffusion velocity of the particle is given by:
\begin{equation}\label{vx}
 v(x) \propto x(1-x).
\end{equation}
In other words, as the particle approaches an endpoint, either 0 or 1, its motion slows down, ensuring a smooth and continuous end to the Brownian motion.

To be consistent with Born's rule \cite{10.1007/BF01397184}, the probability of the particle reaching endpoint 1 should be \(x\), and the probability of reaching endpoint 0 should be \(1-x\).  The Born rule can be reproduced in the following way. When the particle starts at a position \(\xi < \frac{1}{2}\), it has an equal probability of reaching either 0 or \(2\xi\), although the diffusion coefficient \(D(x)\) is not constant, as long as the Brownian motion does not favor one direction over the other.  When the particle starts at a position \(\xi > \frac{1}{2}\), it has an equal probability of reaching either 1 or \(2\xi-1\). By combining these two scenarios, we can study Brownian motion starting at any position \(x\).

\begin{figure}[h]\centering
        \vspace{20pt}
        \includegraphics[width=7cm]{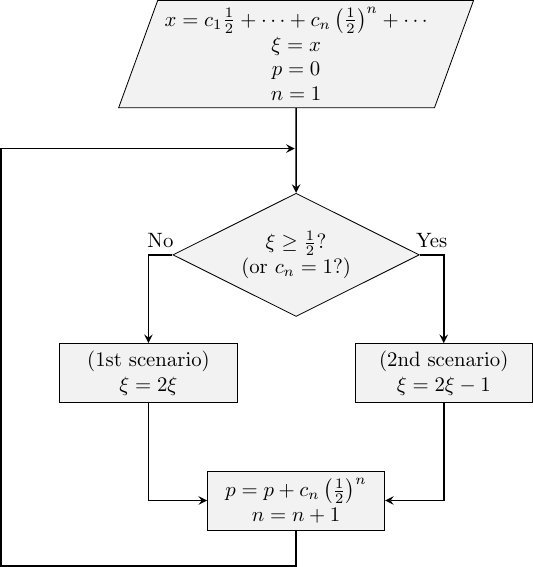}
\caption{
The flowchart depicting the process of a particle undergoing Brownian motion with an initial position \(x\). It shows that the probability \(p\) of the particle reaching endpoint 1 is equal to \(x\), the initial position of the particle.
}\label{prime2}
\end{figure}

To proceed, we expand \(x\) (\(0 \leq x \leq 1\)) as follows:
\begin{equation}
x = c_1\frac{1}{2} + c_2\left(\frac{1}{2}\right)^2 + \cdots + c_n\left(\frac{1}{2}\right)^n + \cdots,
\end{equation}
where each of \(c_1, c_2, \ldots, c_n,\ldots\) is either 0 or 1. Using this representation, we can calculate the probability of the particle reaching endpoint 1 by following the flowchart in Fig. \ref{prime2}. The flowchart demonstrates that the probability of the particle reaching endpoint 1 is \(x\). This method of reproducing Born's rule is similar to the gambler's ruin mechanism \cite{10.1007/BF00726850,shimony_honor_1997}.

\section{Model Validation} 
The impact of vacuum fluctuations on photon absorption by two atoms, as shown in Fig.\ref{atoms}, can be experimentally probed using the Casimir effect\cite{10.1103/PhysRev.73.360,10.1103/PhysRevLett.78.5}. If vacuum fluctuations are crucial in completing photon absorption, this process would be characterized by time-dependence and modulation by fluctuation intensity. The Casimir effect reveals that this intensity can be controlled by adjusting the proximity of parallel conductive plates; reducing the gap weakens the fluctuations. In an adapted setup, positioning these plates around the two atoms from Fig.\ref{atoms} could prolong their entanglement. During this extended entanglement, the atoms have a non-zero probability of simultaneously re-emitting the photon, resulting in an observable interference pattern akin to a double-slit experiment\cite{10.1103/PhysRevLett.70.2359}. The observation window for this pattern, or the time duration over which the interference occurs, can be measured. If vacuum fluctuations are indeed responsible for disrupting the entanglement, then a correlation should exist: the closer the plates, the wider the observation window. This provides a foundation for experimental validation.

Additionally, this experimental setup could further validate our proposed mechanism. Following photon absorption, spontaneous emission, as per our model, would not produce a two-slit interference pattern but rather a single-slit diffraction pattern, attributed to the photon being absorbed by a single atom. Therefore, while observations within the excitation window are expected to show a double-slit pattern, a shift to a single-slit pattern is predicted post-excitation. This transition contrasts with standard quantum mechanics predictions. Drawing from Schrödinger's cat thought experiment~\cite{10.1007/BF01491891}, which posits that unobserved states maintain coherence, interpreting the superposition in Eq.(\ref{3}) as a Schrödinger's cat state implies that coherence would persist through photon absorption and into the spontaneous emission phase, resulting in ongoing two-slit interference. This deviation can be leveraged to test our mechanism by observing whether there is a shift from a two-slit to a single-slit interference pattern, depending on the timing relative to the initiation of photon absorption. For accurate results, it is crucial to ensure that the excited state's lifetime significantly surpasses the duration of excitation.

\section{Three-Atom Photon Absorption}
\begin{figure}[h]
    \centering
    \includegraphics[width=7cm]{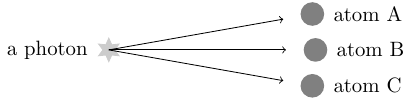}
    \caption{A photon interacting with three atoms, eventually being absorbed by one of them.}
    \label{threeatoms}
\end{figure}

A similar study can be carried out for scenarios in which a single photon is absorbed by more than two atoms. In a three-atom scenario, as depicted in Fig. \ref{threeatoms}, the wave function \(\vert \Psi_{\rm I}\rangle\) assumes the following form when it is stuck:
\begin{equation}
\vert \Psi_{\rm I}\rangle= C_1\vert 1\rangle_A\vert 0\rangle_B\vert 0\rangle_C + C_2\vert 0\rangle_A\vert 1\rangle_B\vert 0\rangle_C + C_3\vert 0\rangle_A\vert 0\rangle_B\vert 1\rangle_C,
\end{equation}
where \(C_1, C_2,\) and \(C_3\) are complex probability amplitudes that satisfy \(|C_1|^2 + |C_2|^2 + |C_3|^2 = 1\). The three atoms are entangled. We define \(x_1\) and \(x_2\) as follows:
\begin{equation}
|C_1|^2=x_1, \quad |C_2|^2=x_2 - x_1 , \quad |C_3|^2=1 - x_2  .
\end{equation}

\begin{figure}[h]
    \centering
    \includegraphics[width=9cm]{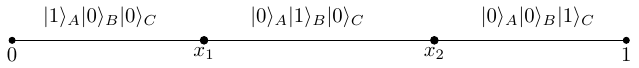}
    \caption{Two particles undergoing Brownian motion on a line of unit length. Upon collision, the two particles merge into a single particle.}
    \label{3parts}
\end{figure}

The disentanglement can be likened to two particles undergoing Brownian motion on the same unit length line, starting at positions \(x_1\) and \(x_2\) respectively, as depicted in Fig. \ref{3parts}. Upon collision, the particles merge. There are three possible outcomes:
\begin{itemize}
    \item Both reach endpoint 1, with probability \(x_1\).
    \item Both reach endpoint 0, with probability \(1 - x_2\).
    \item One at each endpoint, with probability \(x_2 - x_1\).
\end{itemize}
The disentanglement leads to one of the following states: 
\begin{itemize}
    \item \(\vert 1\rangle_A\vert 0\rangle_B\vert 0\rangle_C\), with probability \(x_1\), or $|C_1|^2$.
    \item \(\vert 0\rangle_A\vert 0\rangle_B\vert 1\rangle_C\), with probability \(1 - x_2\), or $|C_3|^2$.
    \item \(\vert 0\rangle_A\vert 1\rangle_B\vert 0\rangle_C\), with probability \(x_2 - x_1\), or $|C_2|^2$.
\end{itemize}
This framework can be generalized to include more atoms by adding additional particles, while still employing the same Brownian motion analogy.

\section{One-Atom Photon Absorption}
A single atom in a superposition of discrete states \(|0\rangle\) and \(|1\rangle\) may also interact with vacuum fluctuations. To illustrate this, consider a scenario with only one atom absorbing a photon. Initially, the system's wave function is \(|\gamma\rangle|0\rangle\), where \(|\gamma\rangle\) denotes the wave function of the photon. Upon interaction, the wave function evolves to:
\begin{equation}
|\Psi_{\text{I}}\rangle = C_1|\gamma\rangle|0\rangle + C_2|1\rangle,
\end{equation}
where \(C_1\) and \(C_2\) satisfy \(|C_1|^2 + |C_2|^2 = 1\). Rewriting this equation yields:
\begin{equation}
|\Psi_{\text{I}}\rangle = (|\gamma\rangle + 1)(C_1|0\rangle + C_2|1\rangle) - C_1|0\rangle - C_2|\gamma\rangle|1\rangle.
\end{equation}

The atom is partially in the superposition state \(C_1|0\rangle + C_2|1\rangle\), enabling it to interact with vacuum fluctuations. This introduces randomness: vacuum fluctuations may occasionally prevent the photon from being absorbed and alter its wave function in such a way that it goes out into a random direction.

This example could be considered a special case of the CSL model and vacuum fluctuations could serve as the random term. However, this random term is likely to be significant only when the atom is in a superposition state.

Extending this example to multi-atom systems, such as the one depicted in Fig.~\ref{atoms}, suggests that vacuum fluctuations should initiate as soon as atoms enter a superposition. The overall wave function then becomes:
\begin{equation}
|\Psi_{\text{I}}\rangle = C_1|\gamma\rangle|0\rangle_A|0\rangle_B + C_2|1\rangle_A|0\rangle_B + C_3|0\rangle_A|1\rangle_B,
\end{equation}
with \(C_1, C_2, C_3\) satisfying \(|C_1|^2 + |C_2|^2 + |C_3|^2 = 1\). 
The Schr\"{o}dinger equation and vacuum fluctuations govern the system's evolution simultaneously as soon as the interaction starts. Specifically, while the Schr\"{o}dinger equation governs the transition from the first term to the second and third terms, vacuum fluctuations cause the coefficients \(C_1\), \(C_2\), and \(C_3\) to fluctuate.

\section{Information and Vacuum Fluctuations}
In the absence of vacuum fluctuations, the complete absorption of a photon by an atom would be impossible. Governed solely by the deterministic Schrödinger equation, this process would require the photon's wave function to align precisely with the atom's discrete eigenstates to facilitate transitions. However, given that the photon's wave function possesses a continuous spatial configuration established prior to its interaction with the atom, there is an inherently near-zero probability of achieving such precise alignment.

Vacuum fluctuations offer a solution to these issues by facilitating the transition from the initial to the final state in two ways, which operate concurrently:
\begin{itemize}
\item The Schrödinger equation maps the initial state—which includes both the photon and the atom in its ground state—to an output that closely approximates one of the atom's excited states. Any deviations manifest as vacuum fluctuations, which eventually decouple from the excited state, preserving unrepresented information from the initial state. It's worth noting that this preserved information is not readily extractable and remains hidden. Additionally, treating the output as a superposition of the excited state and vacuum fluctuations is valid only when all conservation laws are met.
\item Most of the time, the Schrödinger equation maps the initial state to an output that is a general superposition of the ground and excited states. Pre-existing vacuum fluctuations induce Brownian motion in this output state, steering it closer to an eigenstate.
\end{itemize}


Our study suggests that vacuum fluctuations could be central to explaining the randomness observed in the photon absorption process by atoms. These fluctuations, integral to quantum field theory, are essential for accurately calculating phenomena such as the Lamb shift~\cite{10.1103/PhysRev.72.241,10.1103/PhysRev.72.339}. Yet, the deterministic nature of quantum field theory poses an intriguing paradox: how does it give rise to randomness in photon absorption? Quantum field theory's role in time evolution implies that vacuum fluctuations are dynamic, a notion visually exemplified in Quantum Chromodynamics (QCD)\cite{10.1073/pnas.0501642102,zotero-954}. Different from the Lamb shift, photon absorption is a dynamic process and can be influenced by the evolving nature of vacuum fluctuations. Should these fluctuations indeed be dynamic, their specific initial conditions could hold latent information that manifests as apparent randomness during the photon absorption process, as guided by the dynamics of quantum field theory. This phenomenon mirrors the principles of chaos theory\cite{10.1175/1520-0469}, effectively bridging deterministic and probabilistic frameworks in physics.

\section{Summary}
We have investigated scenarios involving the absorption of a single photon by one, two, and three atoms to address several fundamental issues. Firstly, we demonstrate that this process can be continuous, with the Schrödinger equation playing a significant role in its time evolution. Secondly, vacuum fluctuations are shown to contribute to this evolution, introducing randomness. Thirdly, by exploring single-system scenarios instead of ensembles, we successfully replicate the Born rule in this context through a Brownian motion analogy. Fourthly, the challenge posed by the discrete nature of atomic eigenstates implies that external forces must intervene, leading to information loss and the emergence of randomness. Additionally, we propose an experiment to investigate the impact of vacuum fluctuations on two atoms absorbing a single photon in a double-slit setup, with the atoms acting as slits. This experiment aims to ascertain whether the Casimir effect can prolong the timeframe for creating a two-slit interference pattern.

\newgeometry{left=1in, right=1in, top=1in, bottom=1in}                                                       \singlespacing

\vspace{2cm}
\noindent {\bf Acknowledgements} The authors have no relevant financial or non-financial interests to disclose.

\vspace{0.5cm}
\noindent {\bf Data availability} 
All data generated or analysed during this study are included in this published article.
\vspace{2cm}


\begin{thebibliography}{10}
\expandafter\ifx\csname url\endcsname\relax
  \def\url#1{\burl{#1}}\fi
\expandafter\ifx\csname urlprefix\endcsname\relax\def\urlprefix{URL }\fi
\providecommand{\bibinfo}[2]{#2}
\providecommand{\eprint}[2][]{\url{#2}}
\providecommand{\doi}[1]{\url{https://doi.org/#1}}
\bibcommenthead

\bibitem{10.1017/CBO9780511813993}
\bibinfo{author}{Scully, M.~O.} \& \bibinfo{author}{Zubairy, M.~S.}
\newblock \emph{\bibinfo{title}{Quantum {{Optics}}}}
  (\bibinfo{publisher}{{Cambridge University Press}},
  \bibinfo{address}{{Cambridge}}, \bibinfo{year}{1997}).

\bibitem{10.1007/978-1-4757-0576-8}
\bibinfo{author}{Shankar, R.}
\newblock \emph{\bibinfo{title}{Principles of {{Quantum Mechanics}}}}
  (\bibinfo{publisher}{{Springer US}}, \bibinfo{address}{{New York, NY}},
  \bibinfo{year}{1994}).

\bibitem{zotero-937}
\bibinfo{author}{Rudin, W.}
\newblock \emph{\bibinfo{title}{Real and Complex Analysis}}
  \bibinfo{edition}{3rd ed} edn (\bibinfo{publisher}{{McGraw-Hill}},
  \bibinfo{address}{{New York}}, \bibinfo{year}{1987}).

\bibitem{10.1038/121580a0}
\bibinfo{author}{Bohr, N.}
\newblock \bibinfo{title}{The {{Quantum Postulate}} and the {{Recent
  Development}} of {{Atomic Theory1}}}.
\newblock \emph{\bibinfo{journal}{Nature}}
  \textbf{\bibinfo{volume}{121}}~(3050), \bibinfo{pages}{580--590}
  (\bibinfo{year}{1928}).
\newblock \doi{10.1038/121580a0} .

\bibitem{10.1007/BF01397280}
\bibinfo{author}{Heisenberg, W.}
\newblock \bibinfo{title}{{\"Uber den anschaulichen Inhalt der
  quantentheoretischen Kinematik und Mechanik}}.
\newblock \emph{\bibinfo{journal}{Zeitschrift f\"ur Physik}}
  \textbf{\bibinfo{volume}{43}}~(3), \bibinfo{pages}{172--198}
  (\bibinfo{year}{1927}).
\newblock \doi{10.1007/BF01397280} .

\bibitem{10.1007/BF01397184}
\bibinfo{author}{Born, M.}
\newblock \bibinfo{title}{{Quantenmechanik der Sto\ss vorg\"ange}}.
\newblock \emph{\bibinfo{journal}{Zeitschrift f\"ur Physik}}
  \textbf{\bibinfo{volume}{38}}~(11), \bibinfo{pages}{803--827}
  (\bibinfo{year}{1926}).
\newblock \doi{10.1007/BF01397184} .

\bibitem{10.1007/BF01491891}
\bibinfo{author}{Schr{\"o}dinger, E.}
\newblock \bibinfo{title}{{Die gegenw\"artige Situation in der
  Quantenmechanik}}.
\newblock \emph{\bibinfo{journal}{Die Naturwissenschaften}}
  \textbf{\bibinfo{volume}{23}}~(48), \bibinfo{pages}{807--812}
  (\bibinfo{year}{1935}).
\newblock \doi{10.1007/BF01491891} .

\bibitem{10.1103/PhysRev.47.777a}
\bibinfo{author}{Einstein, A.}, \bibinfo{author}{Podolsky, B.} \&
  \bibinfo{author}{Rosen, N.}
\newblock \bibinfo{title}{Can {{Quantum-Mechanical Description}} of {{Physical
  Reality Be Considered Complete}}?}
\newblock \emph{\bibinfo{journal}{Physical Review}}
  \textbf{\bibinfo{volume}{47}}~(10), \bibinfo{pages}{777--780}
  (\bibinfo{year}{1935}).
\newblock \doi{10.1103/PhysRev.47.777} .

\bibitem{10.1103/PhysicsPhysiqueFizika.1.195a}
\bibinfo{author}{Bell, J.~S.}
\newblock \bibinfo{title}{On the {{Einstein Podolsky Rosen}} paradox}.
\newblock \emph{\bibinfo{journal}{Physics Physique Fizika}}
  \textbf{\bibinfo{volume}{1}}~(3), \bibinfo{pages}{195--200}
  (\bibinfo{year}{1964}).
\newblock \doi{10.1103/PhysicsPhysiqueFizika.1.195} .

\bibitem{10.1051/jphysrad:0192700805022500}
\bibinfo{author}{{de Broglie}, L.}
\newblock \bibinfo{title}{{La m\'ecanique ondulatoire et la structure atomique
  de la mati\`ere et du rayonnement}}.
\newblock \emph{\bibinfo{journal}{Journal de Physique et le Radium}}
  \textbf{\bibinfo{volume}{8}}~(5), \bibinfo{pages}{225--241}
  (\bibinfo{year}{1927}).
\newblock \doi{10.1051/jphysrad:0192700805022500} .

\bibitem{10.1103/PhysRev.85.166}
\bibinfo{author}{Bohm, D.}
\newblock \bibinfo{title}{A {{Suggested Interpretation}} of the {{Quantum
  Theory}} in {{Terms}} of "{{Hidden}}" {{Variables}}. {{I}}}.
\newblock \emph{\bibinfo{journal}{Physical Review}}
  \textbf{\bibinfo{volume}{85}}~(2), \bibinfo{pages}{166--179}
  (\bibinfo{year}{1952}).
\newblock \doi{10.1103/PhysRev.85.166} .

\bibitem{zotero-956}
\bibinfo{author}{Bricmont, J.}
\newblock \emph{\bibinfo{title}{Making Sense of Quantum Mechanics}}
  \bibinfo{edition}{Softcover reprint of the hardcover 1st edition} edn
  (\bibinfo{publisher}{{Springer International Publishing}},
  \bibinfo{address}{{Cham}}, \bibinfo{year}{2016}).

\bibitem{10.1103/RevModPhys.29.454}
\bibinfo{author}{Everett, H.}
\newblock \bibinfo{title}{"{{Relative State}}" {{Formulation}} of {{Quantum
  Mechanics}}}.
\newblock \emph{\bibinfo{journal}{Reviews of Modern Physics}}
  \textbf{\bibinfo{volume}{29}}~(3), \bibinfo{pages}{454--462}
  (\bibinfo{year}{1957}).
\newblock \doi{10.1103/RevModPhys.29.454} .

\bibitem{saunders2010many}
\bibinfo{editor}{Saunders, S.}, \bibinfo{editor}{Barrett, J.},
  \bibinfo{editor}{Kent, A.} \& \bibinfo{editor}{Wallace, D.} (eds)
  \emph{\bibinfo{title}{Many Worlds?: {{Everett}}, Quantum Theory, \& Reality}}
   (\bibinfo{publisher}{{Oxford University Press}}, \bibinfo{address}{{Oxford,
  UK}}, \bibinfo{year}{2010}).

\bibitem{10.1007/BF00708656}
\bibinfo{author}{Zeh, H.~D.}
\newblock \bibinfo{title}{On the interpretation of measurement in quantum
  theory}.
\newblock \emph{\bibinfo{journal}{Foundations of Physics}}
  \textbf{\bibinfo{volume}{1}}~(1), \bibinfo{pages}{69--76}
  (\bibinfo{year}{1970}).
\newblock \doi{10.1007/BF00708656} .

\bibitem{10.1103/PhysRevD.24.1516}
\bibinfo{author}{Zurek, W.~H.}
\newblock \bibinfo{title}{Pointer basis of quantum apparatus: {{Into}} what
  mixture does the wave packet collapse?}
\newblock \emph{\bibinfo{journal}{Physical Review D}}
  \textbf{\bibinfo{volume}{24}}~(6), \bibinfo{pages}{1516--1525}
  (\bibinfo{year}{1981}).
\newblock \doi{10.1103/PhysRevD.24.1516} .

\bibitem{joos2003decoherence}
\bibinfo{author}{Joos, E.} \emph{et~al.}
\newblock \emph{\bibinfo{title}{Decoherence and the Appearance of a Classical
  World in Quantum Theory}}  (\bibinfo{publisher}{{Springer}},
  \bibinfo{year}{2003}).

\bibitem{10.1103/RevModPhys.75.715}
\bibinfo{author}{Zurek, W.~H.}
\newblock \bibinfo{title}{Decoherence, einselection, and the quantum origins of
  the classical}.
\newblock \emph{\bibinfo{journal}{Reviews of Modern Physics}}
  \textbf{\bibinfo{volume}{75}}~(3), \bibinfo{pages}{715--775}
  (\bibinfo{year}{2003}).
\newblock \doi{10.1103/RevModPhys.75.715} .

\bibitem{10.1103/RevModPhys.76.1267}
\bibinfo{author}{Schlosshauer, M.}
\newblock \bibinfo{title}{Decoherence, the measurement problem, and
  interpretations of quantum mechanics}.
\newblock \emph{\bibinfo{journal}{Reviews of Modern Physics}}
  \textbf{\bibinfo{volume}{76}}~(4), \bibinfo{pages}{1267--1305}
  (\bibinfo{year}{2005}).
\newblock \doi{10.1103/RevModPhys.76.1267} .

\bibitem{10.1016/j.physrep.2019.10.001}
\bibinfo{author}{Schlosshauer, M.}
\newblock \bibinfo{title}{Quantum decoherence}.
\newblock \emph{\bibinfo{journal}{Physics Reports}}
  \textbf{\bibinfo{volume}{831}}, \bibinfo{pages}{1--57}
  (\bibinfo{year}{2019}).
\newblock \doi{10.1016/j.physrep.2019.10.001} .

\bibitem{10.1103/PhysRevD.34.470}
\bibinfo{author}{Ghirardi, G.~C.}, \bibinfo{author}{Rimini, A.} \&
  \bibinfo{author}{Weber, T.}
\newblock \bibinfo{title}{Unified dynamics for microscopic and macroscopic
  systems}.
\newblock \emph{\bibinfo{journal}{Physical Review D}}
  \textbf{\bibinfo{volume}{34}}~(2), \bibinfo{pages}{470--491}
  (\bibinfo{year}{1986}).
\newblock \doi{10.1103/PhysRevD.34.470} .

\bibitem{10.1103/PhysRevA.40.1165}
\bibinfo{author}{Di{\'o}si, L.}
\newblock \bibinfo{title}{Models for universal reduction of macroscopic quantum
  fluctuations}.
\newblock \emph{\bibinfo{journal}{Physical Review A}}
  \textbf{\bibinfo{volume}{40}}~(3), \bibinfo{pages}{1165--1174}
  (\bibinfo{year}{1989}).
\newblock \doi{10.1103/PhysRevA.40.1165} .

\bibitem{10.1016/S0370-15730300103-0}
\bibinfo{author}{Bassi, A.} \& \bibinfo{author}{Ghirardi, G.}
\newblock \bibinfo{title}{Dynamical reduction models}.
\newblock \emph{\bibinfo{journal}{Physics Reports}}
  \textbf{\bibinfo{volume}{379}}~(5), \bibinfo{pages}{257--426}
  (\bibinfo{year}{2003}).
\newblock \doi{10.1016/S0370-1573(03)00103-0} .

\bibitem{10.1007/978-3-030-31146-9_1}
\bibinfo{author}{Carlesso, M.} \& \bibinfo{author}{Donadi, S.}
\newblock \bibinfo{editor}{Vacchini, B.}, \bibinfo{editor}{Breuer, H.-P.} \&
  \bibinfo{editor}{Bassi, A.} (eds) \emph{\bibinfo{title}{Collapse {{Models}}:
  {{Main Properties}} and the {{State}} of {{Art}} of the {{Experimental
  Tests}}}}.
\newblock (eds \bibinfo{editor}{Vacchini, B.}, \bibinfo{editor}{Breuer, H.-P.}
  \& \bibinfo{editor}{Bassi, A.}) \emph{\bibinfo{booktitle}{Advances in {{Open
  Systems}} and {{Fundamental Tests}} of {{Quantum Mechanics}}}}, Springer
  {{Proceedings}} in {{Physics}}, \bibinfo{pages}{1--13}
  (\bibinfo{publisher}{{Springer International Publishing}},
  \bibinfo{address}{{Cham}}, \bibinfo{year}{2019}).

\bibitem{10.1103/PhysRevA.39.2277}
\bibinfo{author}{Pearle, P.}
\newblock \bibinfo{title}{Combining stochastic dynamical state-vector reduction
  with spontaneous localization}.
\newblock \emph{\bibinfo{journal}{Physical Review A}}
  \textbf{\bibinfo{volume}{39}}~(5), \bibinfo{pages}{2277--2289}
  (\bibinfo{year}{1989}).
\newblock \doi{10.1103/PhysRevA.39.2277} .

\bibitem{10.1103/PhysRevA.42.78}
\bibinfo{author}{Ghirardi, G.~C.}, \bibinfo{author}{Pearle, P.} \&
  \bibinfo{author}{Rimini, A.}
\newblock \bibinfo{title}{Markov processes in {{Hilbert}} space and continuous
  spontaneous localization of systems of identical particles}.
\newblock \emph{\bibinfo{journal}{Physical Review A}}
  \textbf{\bibinfo{volume}{42}}~(1), \bibinfo{pages}{78--89}
  (\bibinfo{year}{1990}).
\newblock \doi{10.1103/PhysRevA.42.78} .

\bibitem{10.1103/RevModPhys.85.471}
\bibinfo{author}{Bassi, A.}, \bibinfo{author}{Lochan, K.},
  \bibinfo{author}{Satin, S.}, \bibinfo{author}{Singh, T.~P.} \&
  \bibinfo{author}{Ulbricht, H.}
\newblock \bibinfo{title}{Models of wave-function collapse, underlying
  theories, and experimental tests}.
\newblock \emph{\bibinfo{journal}{Reviews of Modern Physics}}
  \textbf{\bibinfo{volume}{85}}~(2), \bibinfo{pages}{471--527}
  (\bibinfo{year}{2013}).
\newblock \doi{10.1103/RevModPhys.85.471} .

\bibitem{10.1103/PhysRevD.94.124036}
\bibinfo{author}{Carlesso, M.}, \bibinfo{author}{Bassi, A.},
  \bibinfo{author}{Falferi, P.} \& \bibinfo{author}{Vinante, A.}
\newblock \bibinfo{title}{Experimental bounds on collapse models from
  gravitational wave detectors}.
\newblock \emph{\bibinfo{journal}{Physical Review D}}
  \textbf{\bibinfo{volume}{94}}~(12), \bibinfo{pages}{124036}
  (\bibinfo{year}{2016}).
\newblock \doi{10.1103/PhysRevD.94.124036} .

\bibitem{Einstein1917}
\bibinfo{author}{Einstein, A.}
\newblock \bibinfo{title}{Zur quantentheorie der strahlung}.
\newblock \emph{\bibinfo{journal}{Physikalische Zeitschrift}}
  \textbf{\bibinfo{volume}{18}}, \bibinfo{pages}{121--128}
  (\bibinfo{year}{1917}).
\newblock \doi{10.1017/S0370164600034429} .

\bibitem{10.1007/BF00726850}
\bibinfo{author}{Pearle, P.}
\newblock \bibinfo{title}{Might god toss coins?}
\newblock \emph{\bibinfo{journal}{Foundations of Physics}}
  \textbf{\bibinfo{volume}{12}}~(3), \bibinfo{pages}{249--263}
  (\bibinfo{year}{1982}).
\newblock \doi{10.1007/BF00726850} .

\bibitem{shimony_honor_1997}
\bibinfo{author}{Pearle, P.}
\newblock \bibinfo{title}{ in \textit{Tales and {{Tails}} and {{Stuff}} and
  {{Nonsense}}}} (eds \bibinfo{editor}{Cohen, R.~S.}, \bibinfo{editor}{Horne,
  M.~A.} \& \bibinfo{editor}{Stachel, J.~S.})
  \emph{\bibinfo{booktitle}{Experimental Metaphysics-Quantum Mechanical Studies
  in Honor of Abner Shimony}}, Vol.~\bibinfo{volume}{1}
  \bibinfo{pages}{143--156} (\bibinfo{publisher}{{Kluwer}},
  \bibinfo{address}{{Great Britain}}, \bibinfo{year}{1997}).
\newblock \urlprefix\url{https://arxiv.org/abs/quant-ph/9805050}.

\bibitem{Bohr1913-BOHOTC}
\bibinfo{author}{Bohr, N.}
\newblock \bibinfo{title}{On the constitution of atoms and molecules, part
  {{I}}}.
\newblock \emph{\bibinfo{journal}{Philosophical Magazine}}
  \textbf{\bibinfo{volume}{26}}, \bibinfo{pages}{1--25} (\bibinfo{year}{1913})
  .

\bibitem{zotero-941}
\bibinfo{author}{Griffiths, D.~J.}
\newblock \emph{\bibinfo{title}{Introduction to Quantum Mechanics}}
  \bibinfo{edition}{Second} edn (\bibinfo{publisher}{{Cambridge University
  Press}}, \bibinfo{address}{{Cambridge}}, \bibinfo{year}{2017}).

\bibitem{zotero-932}
\bibinfo{author}{Feynman, R.~P.}
\newblock \emph{\bibinfo{title}{Quantum Electrodynamics}} Advanced Book
  Classics (\bibinfo{publisher}{{CRC Press}}, \bibinfo{address}{{Boca Raton
  London New York}}, \bibinfo{year}{2018}).

\bibitem{10.1103/PhysRevA.11.814}
\bibinfo{author}{Milonni, P.~W.} \& \bibinfo{author}{Smith, W.~A.}
\newblock \bibinfo{title}{Radiation reaction and vacuum fluctuations in
  spontaneous emission}.
\newblock \emph{\bibinfo{journal}{Physical Review A}}
  \textbf{\bibinfo{volume}{11}}~(3), \bibinfo{pages}{814--824}
  (\bibinfo{year}{1975}).
\newblock \doi{10.1103/PhysRevA.11.814} .

\bibitem{10.1119/1.13886}
\bibinfo{author}{Milonni, P.~W.}
\newblock \bibinfo{title}{Why spontaneous emission?}
\newblock \emph{\bibinfo{journal}{American Journal of Physics}}
  \textbf{\bibinfo{volume}{52}}~(4), \bibinfo{pages}{340--343}
  (\bibinfo{year}{1984}).
\newblock \doi{10.1119/1.13886} .

\bibitem{10.1103/PhysRev.73.360}
\bibinfo{author}{Casimir, H. B.~G.} \& \bibinfo{author}{Polder, D.}
\newblock \bibinfo{title}{The {{Influence}} of {{Retardation}} on the
  {{London-van}} der {{Waals Forces}}}.
\newblock \emph{\bibinfo{journal}{Physical Review}}
  \textbf{\bibinfo{volume}{73}}~(4), \bibinfo{pages}{360--372}
  (\bibinfo{year}{1948}).
\newblock \doi{10.1103/PhysRev.73.360} .

\bibitem{10.1103/PhysRevLett.78.5}
\bibinfo{author}{Lamoreaux, S.~K.}
\newblock \bibinfo{title}{Demonstration of the {{Casimir Force}} in the 0.6 to
  6 {$M$}m {{Range}}}.
\newblock \emph{\bibinfo{journal}{Physical Review Letters}}
  \textbf{\bibinfo{volume}{78}}~(1), \bibinfo{pages}{5--8}
  (\bibinfo{year}{1997}).
\newblock \doi{10.1103/PhysRevLett.78.5} .

\bibitem{10.1098/rspa.1913.0040}
\bibinfo{author}{Bragg, W.~H.} \& \bibinfo{author}{Bragg, W.~L.}
\newblock \bibinfo{title}{The reflection of {{X-rays}} by crystals}.
\newblock \emph{\bibinfo{journal}{Proceedings of the Royal Society of London.
  Series A, Containing Papers of a Mathematical and Physical Character}}
  \textbf{\bibinfo{volume}{88}}~(605), \bibinfo{pages}{428--438}
  (\bibinfo{year}{1997}).
\newblock \doi{10.1098/rspa.1913.0040} .

\bibitem{10.1103/PhysRevLett.70.2359}
\bibinfo{author}{Eichmann, U.} \emph{et~al.}
\newblock \bibinfo{title}{Young's interference experiment with light scattered
  from two atoms}.
\newblock \emph{\bibinfo{journal}{Physical Review Letters}}
  \textbf{\bibinfo{volume}{70}}~(16), \bibinfo{pages}{2359--2362}
  (\bibinfo{year}{1993}).
\newblock \doi{10.1103/PhysRevLett.70.2359} .

\bibitem{10.1086/148307}
\bibinfo{author}{Penzias, A.~A.} \& \bibinfo{author}{Wilson, R.~W.}
\newblock \bibinfo{title}{A {{Measurement}} of {{Excess Antenna Temperature}}
  at 4080 {{Mc}}/s.}
\newblock \emph{\bibinfo{journal}{The Astrophysical Journal}}
  \textbf{\bibinfo{volume}{142}}, \bibinfo{pages}{419--421}
  (\bibinfo{year}{1965}).
\newblock \doi{10.1086/148307} .

\bibitem{10.1063/1.1895505}
\bibinfo{author}{H{\"a}nggi, P.} \& \bibinfo{author}{Marchesoni, F.}
\newblock \bibinfo{title}{Introduction: 100years of {{Brownian}} motion}.
\newblock \emph{\bibinfo{journal}{Chaos: An Interdisciplinary Journal of
  Nonlinear Science}} \textbf{\bibinfo{volume}{15}}~(2),
  \bibinfo{pages}{026101} (\bibinfo{year}{2005}).
\newblock \doi{10.1063/1.1895505} .

\bibitem{10.1016/j.physa.2018.10.017}
\bibinfo{author}{Bhattacharyay, A.}
\newblock \bibinfo{title}{Equilibrium of a {{Brownian}} particle with
  coordinate dependent diffusivity and damping: {{Generalized Boltzmann}}
  distribution}.
\newblock \emph{\bibinfo{journal}{Physica A: Statistical Mechanics and its
  Applications}} \textbf{\bibinfo{volume}{515}}, \bibinfo{pages}{665--670}
  (\bibinfo{year}{2019}).
\newblock \doi{10.1016/j.physa.2018.10.017} .

\bibitem{10.1103/PhysRev.72.241}
\bibinfo{author}{Lamb, W.~E.} \& \bibinfo{author}{Retherford, R.~C.}
\newblock \bibinfo{title}{Fine {{Structure}} of the {{Hydrogen Atom}} by a
  {{Microwave Method}}}.
\newblock \emph{\bibinfo{journal}{Physical Review}}
  \textbf{\bibinfo{volume}{72}}~(3), \bibinfo{pages}{241--243}
  (\bibinfo{year}{1947}).
\newblock \doi{10.1103/PhysRev.72.241} .

\bibitem{10.1103/PhysRev.72.339}
\bibinfo{author}{Bethe, H.~A.}
\newblock \bibinfo{title}{The {{Electromagnetic Shift}} of {{Energy Levels}}}.
\newblock \emph{\bibinfo{journal}{Physical Review}}
  \textbf{\bibinfo{volume}{72}}~(4), \bibinfo{pages}{339--341}
  (\bibinfo{year}{1947}).
\newblock \doi{10.1103/PhysRev.72.339} .

\bibitem{10.1073/pnas.0501642102}
\bibinfo{author}{Wilczek, F.}
\newblock \bibinfo{title}{Asymptotic freedom: {{From}} paradox to paradigm}.
\newblock \emph{\bibinfo{journal}{Proceedings of the National Academy of
  Sciences}} \textbf{\bibinfo{volume}{102}}~(24), \bibinfo{pages}{8403--8413}
  (\bibinfo{year}{2005}).
\newblock \doi{10.1073/pnas.0501642102} .

\bibitem{zotero-954}
\bibinfo{author}{{University of Adelaide}}.
\newblock \bibinfo{title}{Visualizations of {{Quantum Chromodynamics}}}.
\newblock
  \urlprefix\url{http://www.physics.adelaide.edu.au/theory/staff/leinweber/VisualQCD/Nobel/}.

\bibitem{10.1175/1520-0469}
\bibinfo{author}{Lorenz, E.~N.}
\newblock \bibinfo{title}{Deterministic {{Nonperiodic Flow}}}.
\newblock \emph{\bibinfo{journal}{Journal of the Atmospheric Sciences}}
  \textbf{\bibinfo{volume}{20}}~(2), \bibinfo{pages}{130--141}
  (\bibinfo{year}{1963}).
\newblock \doi{10.1175/1520-0469} .

\end{thebibliography}

\end{document}